\RequirePackage{ifpdf}
\ifpdf 
\documentclass[pdftex]{sigma}
\else
\documentclass{sigma}
\fi

\begin{document}

\def\omt{\tilde{\omega}}
\def\ti{\tilde}
\def\o{\Omega}
\def\bchi{\bar\chi^i}
\def\In{{\rm Int}}
\def\ba{\bar a}
\def\w{\wedge}
\def\ep{\epsilon}
\def\k{\kappa}
\def\Tr{{\rm Tr}}
\def\ST{{\rm STr}}
\def\ss{\subset}
\def\rn{\vert \alpha\vert^2}
\def\bi{\bibitem}
\def\ot{\oti\def\om{\omega}
\dmes}
\def\bc{{\bf C}}
\def\bz{{\bf Z}}
\def\ptp{\stackrel{\otimes}{,}}
\def\br{{\bf R}}
\def\de{\delta}
 \def\bt{\beta}
 \def\ve{\vert}
\def\al{\alpha}
\def\la{\langle}
\def\ra{\rangle}
\def\Ga{\Gamma}
\def\st{\stackrel{\wedge}{,}}
\def\stv{\stackrel{\wedge}{\vert}}
\def\th{\theta}
\def\lm{\ti\lambda}
\def\U{\Upsilon}
\def\jp{{1\over 2}}
\def\js{{1\over 4}}
\def\d{\partial}
\def\tr{\triangleright}
\def\trl{\triangleleft}
\def\d{\partial}
\def\bq{\}_{P}}
\def\be{\begin{equation}}
\def\ee{\end{equation}}
\def\bea{\begin{eqnarray}}
\def\eea{\end{eqnarray}}
\def\D{{\cal D}}
\def\A{{\cal A}}
\def\G{{\cal G}}
\def\K{{\cal K}}
\def\H{{\cal H}}
\def\P{{\cal P}}
\def\N{{\cal N}}
\def\R{{\cal R}}
\def\B{{\cal B}}
\def\T{{\cal T}}
\def\bT{\bar{\cal T}}
\def\F{{\cal F}}
\def\n{{1\over n}}
\def\si{\sigma}
\def\ta{\tau}
\def\ov{\over}
\def\l{\lambda}
\def\L{\Lambda}
\def\lpb{{\bf \{}}
\def\rpb{{\bf \}}}

\def\pih{\hat{\pi}}

\def\U{\Upsilon}
\def\e{\varepsilon}
\def\bt{\beta}
\def\ga{\gamma}
\def\om{\omega}
\def\be{\begin{equation}}
\def\ee{\end{equation}}
\def\bea{\begin{eqnarray}}
\def\eea{\end{eqnarray}}
\def\D{{\cal D}}
\def\C{{\cal C}}
\def\G{{\cal G}}
\def\H{{\cal H}}
\def\R{{\cal R}}
\def\B{{\cal B}}
\def\K{{\cal K}}
\def \T{{\cal T}}
\def\S{{\cal S}}
\def\bT{\bar{\cal T}}
\def\F{{\cal F}}
\def\n{{1\over n}}
\def\si{\sigma}

\def\ot{\otimes}
\def\l{\lambda}
\def\L{\Lambda}
\def\ve{\vert}
\def\nr{\nabla_R}
\def\nl{\nabla_L}
\def\pih{\hat{\pi}}
 \def\noi{\noindent}
\def\e{\varepsilon}
\def\bt{\beta}
\def\ga{\gamma}

\allowdisplaybreaks

\renewcommand{\PaperNumber}{079}

\FirstPageHeading

\renewcommand{\thefootnote}{$\star$}

\ShortArticleName{$u$-Deformed WZW Model and Its Gauging}

\ArticleName{$\boldsymbol{u}$-Deformed WZW Model and Its
Gauging\footnote{This paper is a contribution to the Proceedings
of the O'Raifertaigh Symposium on Non-Perturbative and Symmetry
Methods in Field Theory
 (June 22--24, 2006, Budapest, Hungary).
The full collection is available at
\href{http://www.emis.de/journals/SIGMA/LOR2006.html}{http://www.emis.de/journals/SIGMA/LOR2006.html}}}

\Author{Ctirad KLIM\v C\'IK} 
\AuthorNameForHeading{C. Klim\v c\'\i k}

\Address{Institute de math\'ematiques de Luminy,
 163, Avenue de Luminy, 13288 Marseille, France}

\Email{\href{mailto:klimcik@iml.univ-mrs.fr}{klimcik@iml.univ-mrs.fr}}

\ArticleDates{Received September 28, 2006; Published online November 13, 2006}

\Abstract{We review the description of  a particular deformation
of the WZW  model. The resulting theory exhibits
 a Poisson--Lie symmetry with  a non-Abelian cosymmetry group and  can be vectorially gauged.}


\Classification{81T40}

\section{Introduction}

 The  theory of  Poisson--Lie symmetric deformations of the standard WZW models \cite{W} was develo\-ped in
\cite{K04,K05,K06} and it is based on the concept of the twisted
Heisenberg double \cite{ST2}. This contribution is a review of a
part of our work~\cite{K06}. It is intended to the attention of
those readers who are interested just in the direct description
and gauging  of one particular example of the Poisson--Lie WZW
deformation  and do not wish to go through the general theory of
the twisted Heisenberg doubles exposed in~\cite{K06}.

\section[$u$-deformed WZW model]{$\boldsymbol{u}$-deformed WZW model}

 $K$ be a   connected simple compact   Lie group and denote by $(\cdot,\cdot)_\K$ the negative-def\/inite
${\rm Ad}$-invariant Killing form on its  Lie algebra $\K$.   Let
$LK$ be the group of smooth maps from a~circle~$S^1$ into $K$ (the
group law is given by pointwise multiplication) and def\/ine a
non-degenerate ${\rm Ad}$-invariant bilinear form
 $(\cdot\vert \cdot)$ on $ L\K \equiv  Lie (LK)$ by the following formula
\begin{gather}
\label{eq1}
 (\al\ve \bt)=\tfrac{1}{2\pi}\int_{-\pi}^\pi d\si (\al(\si),\bt(\si))_{\K}.
 \end{gather}

 Let $P_{\H}:L\K \to \H$ be the orthogonal projector to the Cartan subalgebra $\H$ of $\K$ and
 let $U:\H\to\H$ be a skew-symmetric  linear operator with respect to the inner product $(\cdot,\cdot)_\K$.
 We denote by $u$ the composition $U\circ P_\H$.

{\it The $u$-deformed WZW model is a dynamical system whose phase
space $P$ is
  the direct product
$P=L\K\times LK$,
 its  symplectic form $\om_u$ reads
\begin{gather}\label{eq2}
 \om_u=\tfrac 12 (dJ_L\w\vert r_{LK})-\tfrac 12(dJ_R\w\vert l_{LK}) + \tfrac 12 (u(dJ_L)\w\vert dJ_L)
+\tfrac 12 (u(dJ_R)\w\vert  dJ_R)
\end{gather}
and its Hamiltonian $H$ is given by}
\[
H=-\tfrac{1}{2k}(J_L\vert J_L) -\tfrac{1}{2k}(J_R\vert J_R).
\]
 Here $k$ is a positive integer,  $r_{LK}=dgg^{-1}$ and $l_{LK}=g^{-1}dg$
 stand for the right and the left-invariant Maurer--Cartan forms on the group manifold $LK$ and the
$L\K$-valued functions $J_L$, $J_R$ on $P$ are def\/ined as
\begin{gather}\label{eq3}
 J_L(\chi,g) = \chi, \qquad J_R(\chi,g) = -{\rm Ad}_{g^{-1}}\chi+kg^{-1}\d_\si g, \qquad g\in LK,\quad \chi\in L\K.
\end{gather}
If $U=0$, the $u$-deformed WZW model becomes just the standard WZW
model in the formulation~\cite{Mad,K04}.

 Consider  the standard actions of the loop group $LK$ on the phase space $P$:
\begin{gather*}
h\rhd_L (\chi,g)=   (k\d_\si hh^{-1} +h\chi h^{-1},hg), \qquad h,g\in LK, \quad \chi\in L\K,\\
h\rhd_R (\chi,g) =(\chi,gh^{-1}),\qquad h,g\in LK, \quad \chi\in
L\K.
\end{gather*}
 It was established in \cite{K06} that these actions can inf\/initesimally be expressed via the
 Poisson bivector $\Pi_u$, corresponding  to the symplectic form $\om_u$:
\begin{gather}\label{eq4}
 \xi_Lf=(\Pi_u(df,J_L^*r_B)\vert \xi), \qquad  \xi_Rf=(\Pi_u(df,J_R^*r_B)\vert \xi).
 \end{gather}
 Here $\xi_L$, $\xi_R$ are, respectively, the vector f\/ields on $P$, corresponding to an element $\xi\in L\K$,
 $f$ is a function on $P$,  $J_{L,R}^*r_B$ stand for pull-backs of the Maurer--Cartan form $r_B$
 on a Lie group  $B$. As a set, $B$ is just $L\K$, however, the group
 law is as follows:
\begin{gather}
\chi\bullet \ti\chi =\chi+e^{u(\chi)}\ti\chi e^{-u(\chi)},\qquad
\chi,\ti\chi\in L\K,\qquad \chi^{-1}=-e^{-u(\chi)}\chi
e^{u(\chi)}.\label{eq5}
\end{gather}
 The reader can recognize in \eqref{eq4} the def\/ining relations of the $LK$-Poisson--Lie symmetries
 with the cosymmetry group equal to $B$ (cf. \cite[eq.~(5.30)]{K04}).

 It is insightful to detail the fundamental relations (4)  in the standard
 Cartan  basis $H^{\mu,n}=H^\mu e^{in\si}\in L\K^{\bc}$, $E^{\al,n}=E^\al e^{in\si}\in L\K^\bc$, $n\in{\mathbb Z}$.
 We have
\begin{gather}
 {H_L^{\mu,m}}f=\{f,J_L^{\mu,m}\}_u,\qquad
  {E_L^{\al,n}}f =\{f,J_L^{\al,n}\}_u  +\langle \al,U(H^\mu)\rangle J_L^{\al,n} \{f,J_L^{\mu,0}\}_u,\label{eq6}\\
  {H_R^{\mu,m}}f=\{f,J_R^{\mu,m}\}_u, \qquad
  {E_R^{\al,n}}f =\{f,J_R^{\al,n}\}_u  +\langle \al,U(H^\mu)\rangle J_R^{\al,n} \{f,J_R^{\mu,0}\}_u,\label{eq7}
  \end{gather}
 where
 \[
 J_{L,R}^{\al,n}\equiv (J_{L,R}\ve E^\al e^{in\si}),\qquad J_{L,R}^{\mu,n}\equiv   (J_{L,R}\ve H^\mu e^{in\si}).
 \]
 For completeness, note that
 $E^\al$ are the step generators of the complexif\/ied Lie algebra $\K^\bc$
 and~$H^\mu$ are the  orthonormalized  generators of the Cartan subalgebra $\H^\bc$:
\begin{gather*}
[H^\mu,E^\al]=\langle \al,H^\mu\rangle E^\al,\qquad
[E^\al,E^{-\al}]=\al^\vee ,
\qquad [E^\al,E^\bt]=c^{\al\bt}E^{\al+\bt},\\
(H^\mu,H^\nu)_\K=\delta^{\mu\nu}, \qquad
(E^{\al},E^{-\al})_{\K^\bc}=\tfrac{2}{\ve \al\ve ^2}, \qquad
(E^\al)^\dagger=E^{-\al},\qquad (H^\mu)^\dagger = H^\mu.
\end{gather*}
 The coroot $\al^\vee$ is def\/ined as
\[
\al^\vee =\tfrac{2}{\ve \al\ve ^2}\langle \al,H^\mu\rangle H^\mu.
\]
  We observe, that the actions $\rhd_{L,R}$ are not Hamiltonian, unless $u=0$. This suggests that
  the current algebra brackets cannot be the same as they are in the non-deformed WZW model.
  Indeed, $u$-corrections are present and  we underline them for the better orientation of the reader:
\begin{gather}
\{J_L^{\mu,m}, J_L^{\nu,n}\}_u=
{k\delta^{\mu\nu}in\delta_{m+n,0}},\qquad
    \{J_L^{\mu,m}, J_L^{\al,n}\}_u=\langle \al,H^\mu\rangle J_L^{\al,n+m},\nonumber\\
\{J_L^{\al,m}, J_L^{-\al,n}\}_u=\tfrac{2}{\ve \al\ve ^2}
\bigl(\langle\al,H^\mu\rangle J_L^{\mu,n+m}
{+ikn\delta_{m+n,0}}\bigr),\nonumber\\
\{J_L^{\al,m},
J_L^{\bt,n}\}_u=c^{\al\bt}J_L^{\al+\bt,m+n}\underline{
-\langle\al,U(H^\mu)\rangle
\langle\bt, H^\mu\rangle   J_L^{\al,m}J_L^{\bt,n}}, \label{eq8}\\
\{J_R^{\mu,m}, J_R^{\nu,n}\}_u=
-{k\delta^{\mu\nu}in\delta_{m+n,0}},\qquad
    \{J_R^{\mu,m}, J_R^{\al,n}\}_u=\langle\al,H^\mu\rangle J_R^{\al,n+m},\nonumber\\
\{J_R^{\al,m}, J_R^{-\al,n}\}_u=\tfrac{2}{\ve \al\ve ^2}
\bigl(\langle\al,H^\mu\rangle
J_R^{\mu,n+m} {-ikn\delta_{m+n,0}}\bigr),\nonumber\\
 \{J_R^{\al,m}, J_R^{\bt,n}\}_u=c^{\al\bt}J_R^{\al+\bt,m+n}\underline{  -\langle\al,U(H^\mu)\rangle
 \langle \bt, H^\mu\rangle   J_R^{\al,m}J_R^{\bt,n}}, \label{eq9}\\
\{J_L,J_R\}_u=0.\label{eq10}
\end{gather}
                Note that the brackets of the left currents dif\/fer from those of the right currents just by the
                sign  in front of the parameter $k$.

The relations \eqref{eq6}, \eqref{eq7} and
\eqref{eq8}--\eqref{eq10} almost determine  the Poisson bracket
$\{\cdot,\cdot\}_u$, corresponding to the symplectic form $\om_u$.
The remaining relation,  which completes the description of
$\{\cdot,\cdot\}_u$, is as follows:
\begin{gather}
 \{\phi,\psi\}_u=U(H^{\mu,0})_L\phi \ H_L^{\mu,0}\psi -H_R^{\mu,0}\phi  \ U(H^{\mu,0})_R\psi.\label{eq11}
\end{gather}
Here $\phi$, $\psi$ are functions on $P$ which depend only on $LK$
but not on $L\K$.

\section{Symplectic reduction}
  The symplectic reduction of a dynamical system $(P,\om)$
consists in singling out a particular set of observables
$\phi_i\in {\rm Fun}(P)$ called f\/irst class  constraints. One
just requires  from $\phi_i$ that on the common locus $L$, where
all $\phi_i$ vanish,
 also all  Poisson brackets  $\{\phi_i,\phi_j\}$ vanish. This requirement and the Frobenius theorem guarantee
that the kernels of the restriction of the  symplectic form $\om$
to $L$
 form an integrable distribution on $L$.  Under certain conditions, the set of integrated surfaces
of this distribution is itself a manifold $P_r$ which is called
the reduced symplectic manifold. The reduced symplectic form
$\om_r$ on $P_r$ is uniquely f\/ixed by  a condition that the
pull-back of $\om_r$ to $L$ coincides with the restriction of the
symplectic form $\om$ to $L$.

In many interesting situations, the integrated surfaces of the
integrable distribution can be naturally identif\/ied with orbits
of a Lie group acting on $L$. This is the reason why  the
symplectic reduction is sometimes called the gauging of that Lie
group action.  As an warm-up example, let us f\/irst perform the
(vectorial) gauging of the standard WZW model corresponding to the
choice $u=0$ in the formula \eqref{eq2}.

 Let $\U$ be a subset of the set of all positive roots of the Lie algebra $\K^\bc$
and suppose that
 the complex vector space $\S^\bc$
\[
\S^\bc={\rm Span}\{E^\ga,E^{-\ga}, [E^\ga,E^{-\ga}]\}, \qquad
\ga\in\U
\]
  is the Lie subalgebra of $\K^\bc$ (as an example take
 the block diagonal embedding
 of $sl_3$ in $sl_4$). The complex Lie algebra $\S^\bc$ has a natural compact real
form $\S$ consisting
 of the anti-Hermitean elements of $\S^\bc$.  Consider the corresponding compact
semi-simple group $S$ and view it as the  subgroup of $K$.

 For the f\/irst class constraints, we take
\begin{gather}
\label{eq12}
 \phi^{\ga,n}\equiv J_L^{\ga,n}+ J_R^{\ga,n},\qquad \phi^{\nu,n}
\equiv J_L^{\nu,n}+J_L^{\nu,n},
\end{gather}
where $\ga\in \pm\U$ and $\nu$ is such that $H^\nu$ is in the
Cartan subalgebra $\H_S$ of $\S$. For  $u=0$, we obtain
\begin{gather*}
\{\phi^{\mu,m}, \phi^{\nu,n}\}_{u=0}= 0,\qquad
    \{\phi^{\mu,m}, \phi^{\al,n}\}_{u=0}=\langle \al,H^\mu\rangle  \phi^{\al,n+m},\\
\{\phi^{\al,m}, \phi^{-\al,n}\}_{u=0}=\tfrac{2}{\ve \al\ve ^2}
\langle \al,H^\mu\rangle\phi^{\mu,n+m},\qquad
                 \{\phi^{\al,m}, \phi^{\bt,n}\}_{u=0}=c^{\al\bt}\phi^{\al+\bt,m+n}.
\end{gather*}
We immediately observe that the Poisson brackets of the f\/irst
class constraints vanish on the common locus $L=\{p\in P;
\phi^{\ga,n}(p)=0, \phi^{\nu,n}(p)=0\}$, therefore the symplectic
reduction can be performed. As the result of analysis, it turns
out that the integrated surfaces of the integrable distribution
are given as the orbits of the following action of the loop group
$LS$ on $L$:
\[
s\rhd (\chi,g)=   \big(k\d_\si ss^{-1} +s\chi
s^{-1},sgs^{-1}\big), \qquad s\in LS,\quad (\chi,g)\in L.
\]

 If $u\neq 0$, the Poisson brackets of the constraints \eqref{eq12} do not vanish on the
common locus $L$ and, therefore, they cannot serve as the base for
a symplectic reduction. It is not dif\/f\/icult to f\/ind a  way
out from the trouble, however. For that, we take an inspiration
from the case $u=0$ where    the sum of the left and right
currents can be interpreted as the product in the Abelian
cosymmetry group $L\K$ (the group multiplication is the addition
in the vector space $L\K$). Thus, for $u\neq 0$, it looks
plausible to use the product \eqref{eq5} in the non-Abelian
cosymmetry group $B$. This gives the following constraints:
\begin{gather*}
 \phi_u^{\ga,n}\equiv (J_L\bullet J_R\ve E^{\ga,n})=
J_L^{\ga,n}+e^{-\langle \ga,U(H^\nu)\rangle J_L^{\nu,0}}J_R^{\ga,n},\\
\phi_u^{\nu,n} \equiv (J_L\bullet J_R\ve
H^{\nu,n})=J_L^{\nu,n}+J_L^{\nu,n},
\end{gather*}
where, again,  $\ga\in \pm\U$ and $\nu$ is such that $H^\nu$ is in
$\H_S$.
  Suppose, moreover,  that  it holds  for all $\ga\in\U$:
\[
(\ga\circ U)(\H_S^\perp)=0,
\]
where the subscript $\perp$ stands for the orthogonal complement
with respect to the restriction of   the Killing--Cartan form
$(\cdot,\cdot)_\K$ to $\H$. Then the Poisson brackets of the
constraints $\phi_u$
 vanish on the common locus $L_u=\{p\in P; \phi_u^{\ga,n}(p)=0,
\phi_u^{\nu,n}(p)=0\}$, as it is obvious from the following
explicit formulas:
\begin{gather*}
\{\phi_u^{\mu,m}, \phi_u^{\nu,n}\}_u=0,\qquad
\{\phi_u^{\mu,m}, \phi_u^{\al,n}\}_u=\langle\al,H^\mu\rangle \phi_u^{\al,n+m},\\
\{\phi_u^{\al,m}, \phi_u^{-\al,n}\}_u=\tfrac{2}{\ve \al\ve ^2}
\langle\al,H^\mu\rangle\phi_u^{\mu,n+m},\\
 \{\phi_u^{\al,m}, \phi_u^{\bt,n}\}_u=c^{\al\bt}\phi_u^{\al+\bt,m+n}
   -\langle \al,U(H^\mu)\rangle \langle \bt, H^\mu\rangle   \phi_u^{\al,m}\phi_u^{\bt,n}.
\end{gather*}
The symplectic reduction now can be performed and the question
arises whether we can identify the orbits of a $LS$ action on
$L_u$ which would coincide with the integrated surfaces of the
integrable distribution.  The answer is af\/f\/irmative \cite{K06}
and  it reads:
\[
s\tr_u (\chi, g)=\big(s\chi s^{-1} +k \partial_\si ss^{-1},
sgs_L^{-1}\big),
 \qquad s_L=e^{-u(sJ_Ls^{-1}+\k \partial ss^{-1})}se^{u(J_L)}.
\]
 We conclude that the reduced symplectic manifold
$P_{ru}$ can be identif\/ied with the coset space $L_u/LS$.

\section{Outlook}
 We believe that
the $u$-deformed WZW model may become a useful laboratory for the
study of possible generalizations of the standard axioms of the
conformal f\/ield theory in two dimensions.

\LastPageEnding
\end{document}